LA-UR-02-3566

*Approved for public release; distribution is unlimited.*

*Title:*

**Induced radioactivity problem for high-power heavy-ion accelerators - Experimental investigation and longtime predictions**

*Author(s):*

DA. Fertman, I. Bakhmetjev, V. Batyaev, N. Borisenko, A. Cherkasov, A. Golubev, A. Kantsyrev, E. Karpikhin, A. Koldobsky, K. Lipatov, R. Mulambetov, S. Mulambetova, Yu. Nekrasov, M. Prokouronov, I. Roudskoy, B. Sharkov, G. Smirnov, Yu. Titarenko, V. Turtikov, V. Zhivun *Institute for Theoretical and Experimental Physics, 117218 Moscow, Russia* G. Fehrenbacher, R.W. Hasse, D.H.H. Hoffmann, I. Hofmann, E. Mustafin, K. Weyrich, J. Wieser *GSI, D-64291Darmstadt, Germany* S. Mashnik *Nuclear Physics Group T-16, Theoretical Division, MS B283, Los Alamos National Laboratory, Los Alamos, NM 87545, USA* V. Barashenkov *Joint Institute for Nuclear Research, 141980 Dubna, Russia* K. Gudima, *Institute of Applied Physics, Academy of Science of Moldova, Kishinev, MD-2028, Moldova*

*Submitted to:*

http://lib-www.lanl.gov/cgi-bin/getfile?00818851.pdf

# Induced radioactivity problem for high-power heavy-ion accelerators - Experimental investigation and longtime predictions

A. Fertman[1], I. Bakhmetjev, V. Batyaev, N. Borisenko, A. Cherkasov, A. Golubev,

A. Kantsyrev, E. Karpikhin, A. Koldobsky, K. Lipatov, R. Mulambetov, S. Mulambetova,

Yu. Nekrasov, M. Prokouronov, I. Roudskoy, B. Sharkov, G. Smirnov, Yu. Titarenko,

V. Turtikov, V. Zhivun

*Institute for Theoretical and Experimental Physics, 117218 Moscow, Russia*

G. Fehrenbacher, R.W. Hasse, D.H.H. Hoffmann, I. Hofmann, E. Mustafin, K. Weyrich,

J. Wieser

*GSI, D-64291Darmstadt, Germany*

S. Mashnik

*Nuclear Physics Group T-16, Theoretical Division, MS B283,*

*Los Alamos National Laboratory, Los Alamos, NM 87545, USA*

V. Barashenkov

*Joint Institute for Nuclear Research, 141980 Dubna, Russia*

K. Gudima

*Institute of Applied Physics, Academy of Science of Moldova, Kishinev, MD-2028, Moldova,*

Corresponding author: Alexander Fertman, Alexander.Fertman@itep.ru, (007095)1299689

**Induced radioactivity problem for accelerators**

**Pages-11**

**Figures-6**

**Tables-0**

## *1. Introduction*

In recent years, several laboratories have started building new high-power heavy-ion accelerators. Among the important problems that have to be solved during the design stage concerns the radiation protection of the accelerator. Three main topics must be studied before the high-power heavy-ion facilities are put into full-scale operations, namely,

*1. Radionuclide production in the components of beam lines;*

*2. Long-term prediction of radioactive inventory around the beam;*

*3. Hazard due to secondary neutrons of heavy-ion beam facilities.*

This work deals with the first and second topics. In present-day heavy-ion accelerators, the activation of structural elements is not a serious problem at relatively low intensities. However, low ion intensities are sufficient to permit measuring the absolute values of dose rates and the cross sections for residual nuclide production, bearing in mind that the database of induced radioactivity for different materials is still poor.

Thus, experimental data, which can be obtained with the present-day facilities (SIS-18, Darmstadt; TWAC - ITEP, Moscow) are important for determining the induced radioactivity in high-intensity heavy-ion accelerators. In November 2000, the decision was made to study the radiation safety aspects for new high-power accelerator facilities.

In view of the above, two experiments were carried out:

- Measurements of activation and dose rates of thick copper targets irradiated with carbon ions of 0.1 GeV/u energy (SIS-18 facility);
- Measurements of residual nuclide production cross sections for thin copper and cobalt targets irradiated with carbon ions of 0.2 GeV/u energy (TWAC - ITEP facility).

The residual nuclide production cross sections obtained in the latter run using the recently constructed TWAC facility can be used in two ways, namely, (1) as data for direct determination of activation and dose loads during accelerator operations and long after its shutdown and (2) as data for verifying different models of nucleus-nucleus interactions. In this work, two models were tested by simulating the production of the measured residual nuclides and subsequently comparing them with experimental data: (1) the Dubna version of the cascade model of nucleus-nucleus interactions CASCADE [Barashenkov, (2000)] that includes also evaporation of light particles up to alphas from excited residual nuclei after the cascade stage of reactions, and (2) the Los Alamos Quark Gluon String Model (LAQGSM) [ Mashnik, (to be published)] merged with Furihata's GEM2 code based on the Generalized Evaporation Model (GEM), LAQGSM+GEM2. After the cascade stage of a reaction, the LAQGSM+GEM2 code considers a possible pre-equilibrium emission of up to 29 different nucleons, complex particles, and light fragments (Z<7), evaporation of up to 66 different particles and light fragments (Z<13), fission of heavy compound nuclei, Fermi break-up of light excited nuclei, and coalescence of complex particles from fast nucleons emitted during the cascade stage of a reaction.

***2. Activation and dose rate measurements at GSI***

During the first test, the radionuclides produced were identified using a γ-spectrometer with a high-purity Ge detector. The spectrometer permits measuring the γ spectra up to 2 MeV. The produced radionuclides were identified by their characteristic γ line energies. The 50-mm diameter, 10-mm thick cylindrical targets were mounted and fixed on a support, which permitted alignment of the targets and the beam axis during the irradiation. The carbon ion range in copper is approximately 4.2 mm (the real energy of the carbon ions in our case is ~97 MeV/u). Thus, the beam was completely stopped in the target. The beam delivery system at Cave A permits uniform irradiation of the targets, or a part of them, with different intensities up to $6 \cdot 10^8$ particles per pulse. For our low-dose measurements the irradiated targets were transported manually to the γ-spectrometer. Three copper samples were irradiated by $5 \cdot 10^9$ ion/cm$^2$ ($t_i$ = 1460 s). Since only a test experiment was carried out, the maximum cooling time for recording the spectra was 10 hours. The dose rates from the samples measured by the conventional dosimeter were: 10 μSv/h on the back surface and 60 μSv/h on the front surface of the target. The γ–spectra of the copper cylinder irradiated by $5 \cdot 10^9$ ion/cm$^2$ (see Fig. 1) were used to predict the activity for the nuclides identified (see Fig. 2). As seen, after 20 days only $^{51}$Cr, $^{52}$Mn, $^{54}$Mn, $^{56}$Co and $^{58}$Co nuclides make a contribution to the dose rate of the target.

Figure 1, Figure 2

***3. Residual nuclides production cross-sections measurements at ITEP***

The cross-sections for residual nuclide production in thin $^{nat}$Cu and $^{59}$Co targets (10.5 mm diameter, 0.25 mm thickness) were measured using a γ-spectrometer without preliminary chemical separation of the nuclides produced. In the experiment, the irradiated targets were placed normally to the carbon-ion beam for an exposure time of 3 hours. The total carbon fluence was determined using a calibrated current transformer and $^{nat}$Cu($^{12}$C,X)$^{24}$Na monitor reaction. The values obtained are $8.2 \cdot 10^{10}$ $^{12}$C/cm$^2$ and $8.4 \cdot 10^{10}$ $^{12}$C/cm$^2$, respectively. The 10 mb value of monitor reaction cross-section for 200 MeV/u energy was taken from [Polanski, (1991)]. In 10 minutes after the irradiation, the experimental targets were measured using a γ-spectrometer with a Ge detector (resolution of 1.8 keV for the 1332 keV $^{60}$Co γ line). In total, 18 spectra were measured for each of the targets within a month after irradiation. The spectrum measurement time varied from 300 s up to 54,000 s. The GENIE2000 code was used to process the measured γ spectra. After all the measured γ spectra were processed, the SIGMA code and PCNUDAT database were used to identify radionuclides.

Examples of some of decay curves for radionuclides identified in the irradiated Cu and Co targets are shown in Fig. 3 (the total number of the nuclides identified is 26 in the Cu target and 26 in the Co target). The decay curve parameters found by the least-squares method were used to determine the independent and cumulative radionuclide yields [Titarenko, (2002)].

The resultant yields were simulated by the CASCADE and LAQGSM codes. The mean-square deviation factor <F> was chosen to be the simulation-experiment agreement criterion. The values of the criterion for either of the codes can be seen in Fig. 4. The comparison between them has shown that the LAQGSM code is preferable for this case.

Figure 3, Figure 4

The simulation results were used to make long-term predictions of the activity behavior of long-lived radionuclides for a period of $\sim 10^5$ years after irradiation (see Figs. 5 and 6). The greatest difference in the activities obtained by the CASCADE and LAQGSM codes occurs within the cooling times of 50-200 years and over 2000 years.

### *4. Conclusion*

The $\sim 30\,\%$ disagreement between the directly and indirectly measured radiation characteristics of the irradiated thick targets has occurred because the $\gamma$ lines of the radionuclides with insufficiently high production cross sections (yields) are impossible to single out in the $\gamma$ spectra.

The mean square deviation factor that we found between experiment and simulation is ranging from 2 to 3. The yields of some radionuclides ($^{24}$Na, $^{47}$Ca) are predicted much worse by both codes. Therefore, further improvement of the simulation codes necessitates persistent buildup of the relevant experimental data (the thick target experiments included).

Figure 5, Figure 6

### *Acknowledgment*

The work has been partially supported by the U.S. Department of Energy and by CRDF Grant No. MP2-3025.

**Induced radioactivity problem for accelerators**

Figure 1. Spectra of the Cu target irradiated uniformly by 100 MeV/u carbon ion beam up to the dose $5*10^9$ ions/$cm^2$.

Figure 2. Activity predictions for the Cu target irradiated by 100 MeV/u carbon ion beam up to the dose $5*10^9$ ions/$cm^2$

Figure 3. Typical examples of measured counting rates and the fitted decay curves for the Co target (the left-hand panel) and for the Cu target (the right-hand panel).

Figure 4. Detailed comparison between experimental and simulated cross sections of radioactive reaction products in $^{59}$Co and $^{nat}$Cu induced by 0.2 GeV/u $^{12}$C. The cumulative cross sections are labeled with a “cum” when the respective independent cross sections are also shown.

Figure 5. The LAQGSM-simulated activities of long-lived $^{nat}$Cu($^{12}$C,X) reaction products.

Figure 6. The LAQGSM- and CASCADE-simulated total activities of long-lived $^{nat}$Cu($^{12}$C,X) and $^{59}$Co($^{12}$C,X) reaction products.

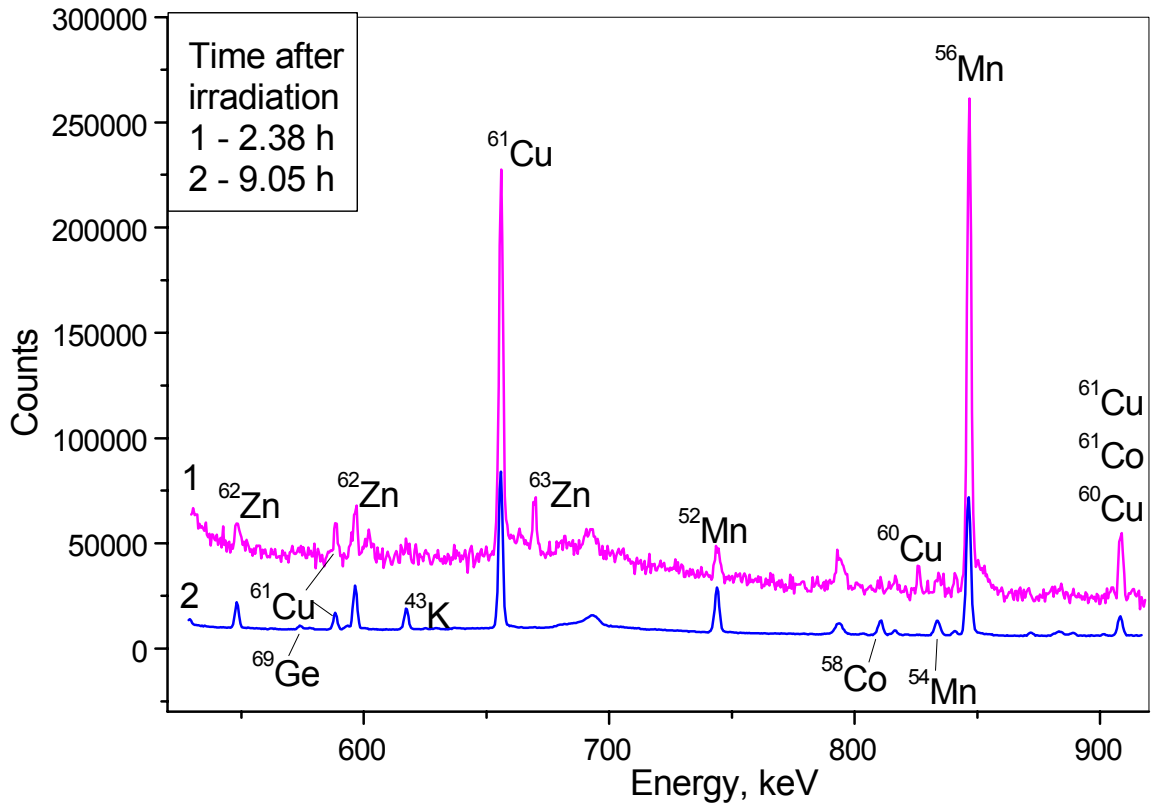

Figure 1.

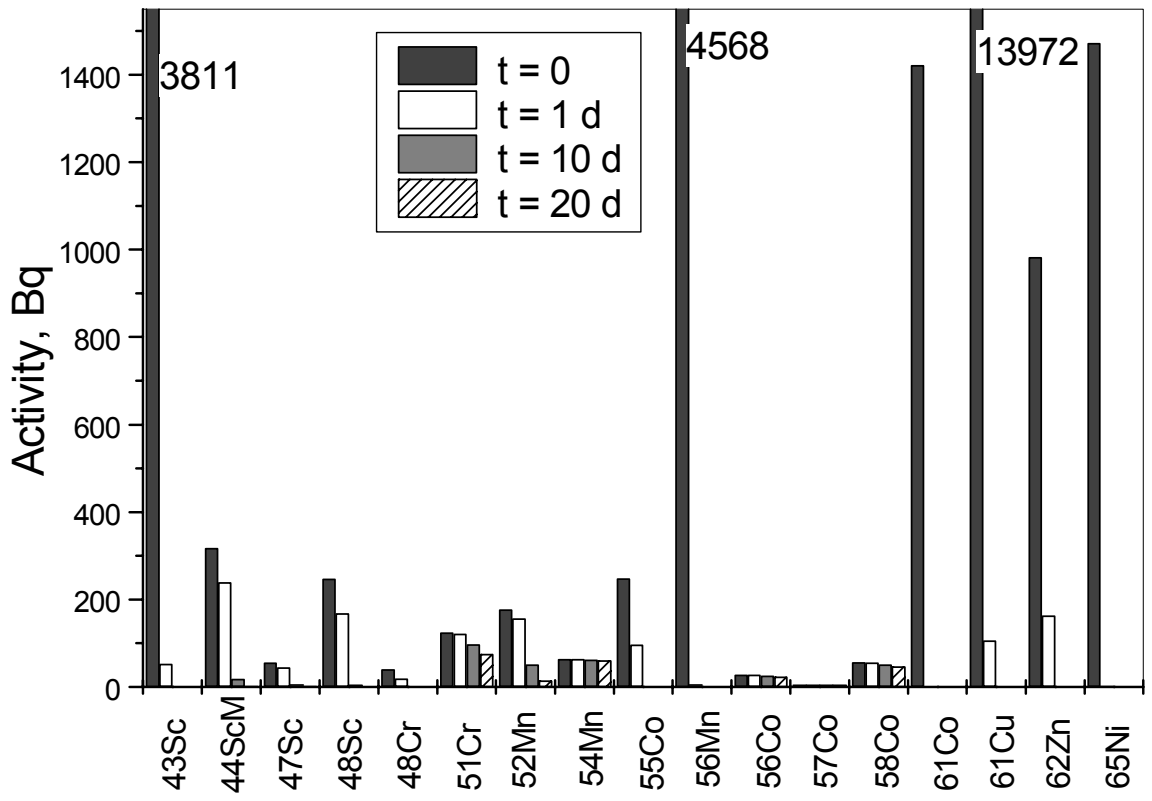

Figure 2.

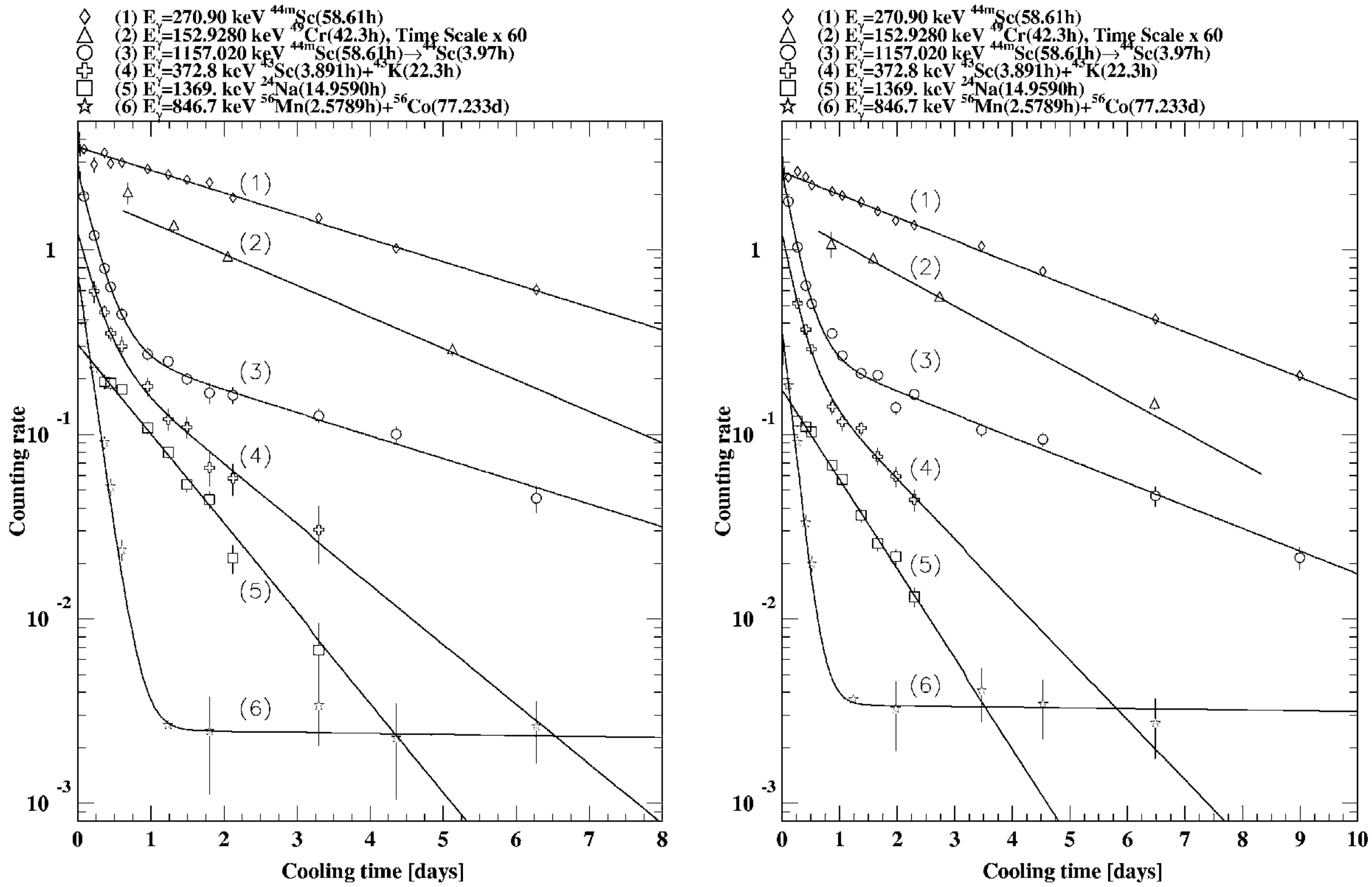
(1) Eγ=270.90 keV 44mSc(58.61h)
(2) Eγ=152.9280 keV 49Cr(42.3h), Time Scale x 60
(3) Eγ=1157.020 keV 44mSc(58.61h)→44Sc(3.97h)
(4) Eγ=372.8 keV 43Sc(3.891h)+43K(22.3h)
(5) Eγ=1369. keV 24Na(14.9590h)
(6) Eγ=846.7 keV 56Mn(2.5789h)+56Co(77.233d)
Counting rate
Cooling time [days]
(1)
(2)
(3)
(4)
(5)
(6)

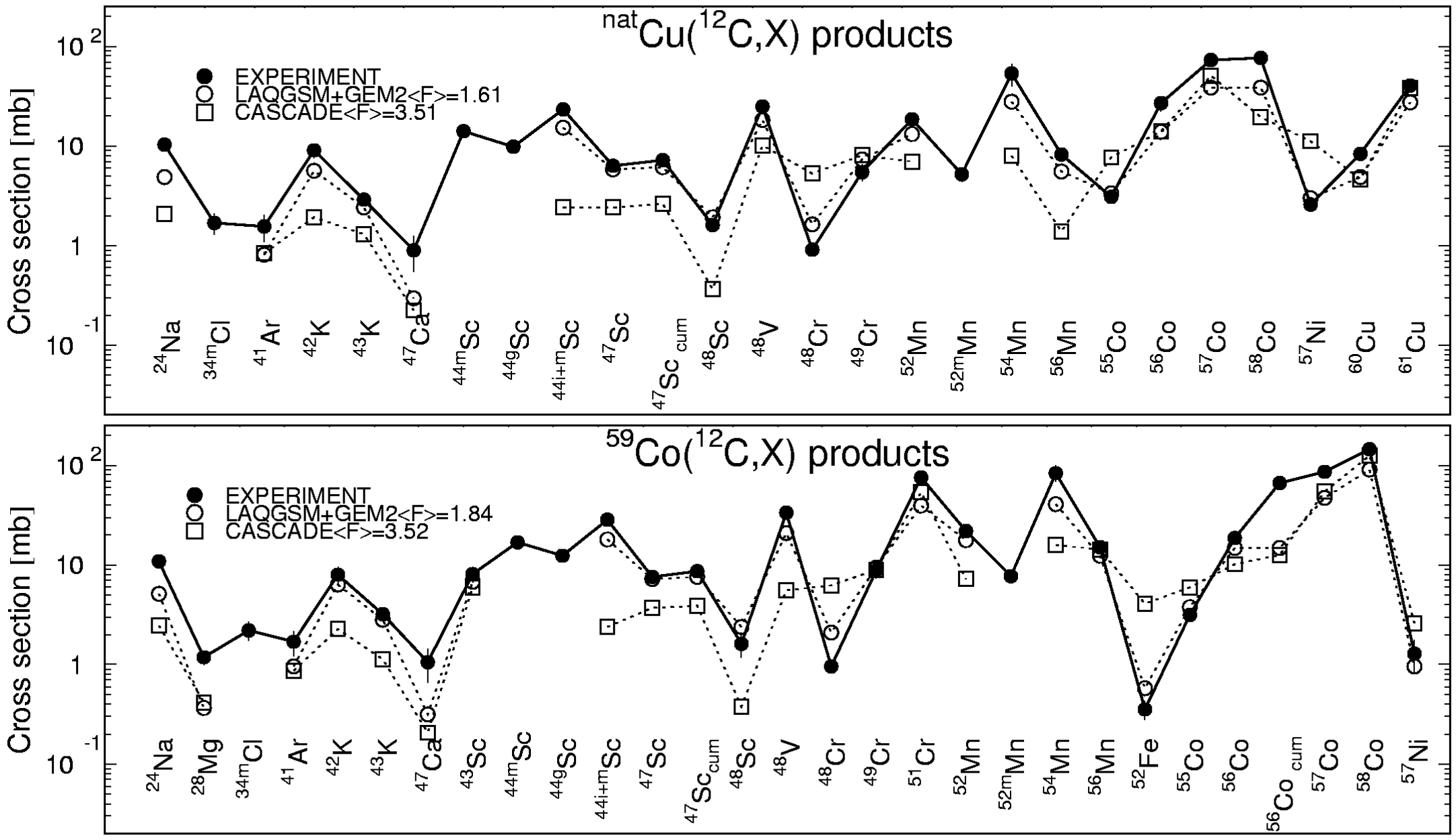

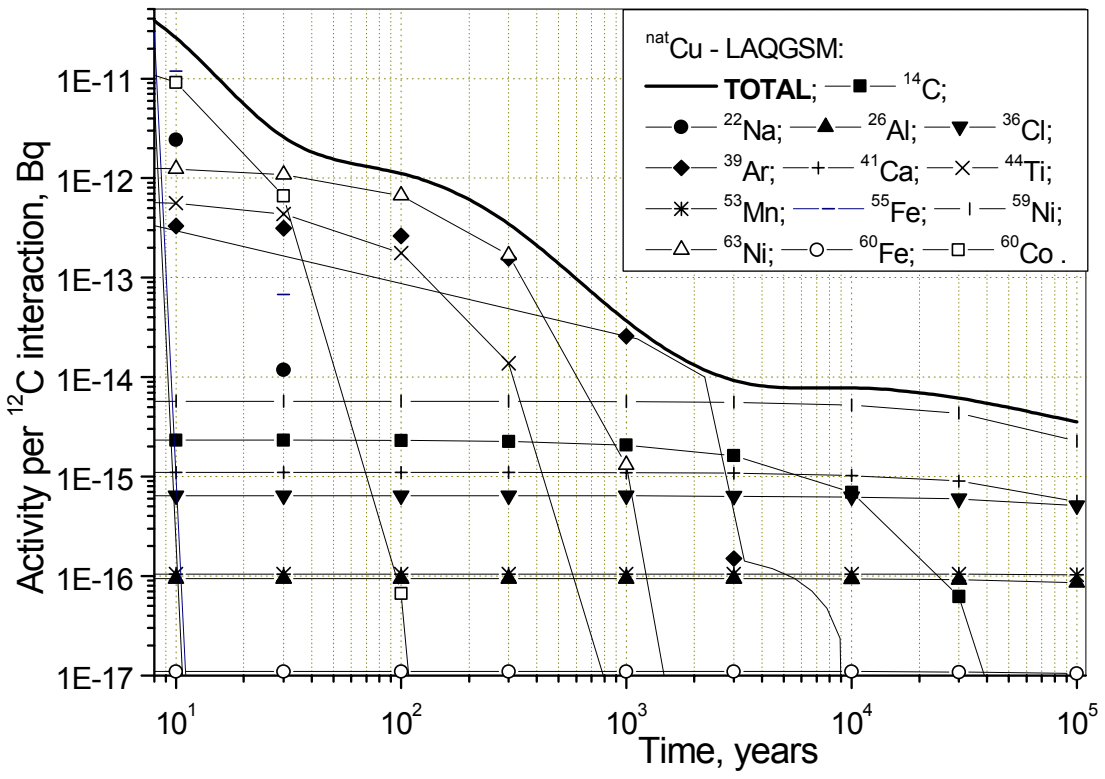

natCu - LAQGSM:
TOTAL; 14C;
22Na; 26Al; 36Cl;
39Ar; 41Ca; 44Ti;
53Mn; 55Fe; 59Ni;
63Ni; 60Fe; 60Co .
Activity per 12C interaction, Bq
Time, years

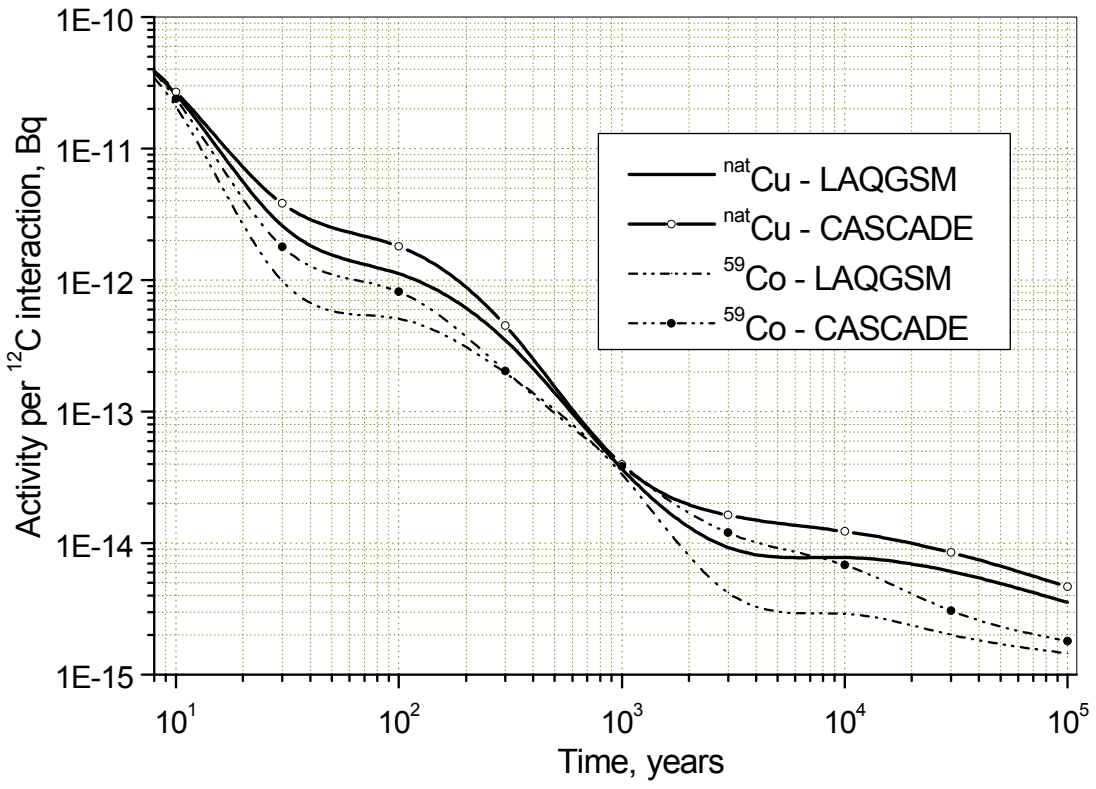